\documentclass[preprints,article,submit,pdftex,moreauthors]{Definitions/mdpi} 

\firstpage{1} 
\pubvolume{1}
\issuenum{1}
\articlenumber{0}
\pubyear{2025}
\copyrightyear{2025}
\datereceived{ } 
\daterevised{ } 
\dateaccepted{ } 
\datepublished{ } 
\hreflink{https://doi.org/}

\Title{Probing Equatorial Ionospheric TEC at Sub-GHz Frequencies with Wide-Band (B4) uGMRT Interferometric Data}

\Author{Dipanjan Banerjee $^{1,\ddagger}$*, Abhik Ghosh $^{1,\ddagger}$*, Sushanta K Mondal $^{2}$, Parimal Ghosh $^{1}$}

\address{$^{1}$ \quad Department of Physics, Banwarilal Bhalotia College, Asansol, West Bengal-713303, India\\
$^{2}$ \quad Department of Physics, Sidho-Kanho-Birsha University, Ranchi Road, Purulia 723104, India}

\corres{Correspondence: DB: dipanjanbanerjee20@gmail.com, AG: abhik.physicist@gmail.com}

\secondnote{These authors contributed equally to this work.}

\abstract{
Phase stability at low radio frequencies is severely impacted by ionospheric propagation delays. Radio interferometers such as the Giant Metrewave Radio Telescope (GMRT) are capable of detecting changes in the ionosphere’s total electron content (TEC) over larger spatial scales and with greater sensitivity compared to conventional tools like the Global Navigation Satellite System (GNSS). Thanks to its unique design—featuring both a dense central array and long outer arms—and its strategic location, the GMRT is particularly well-suited for studying the sensitive ionospheric region located between the northern peak of the Equatorial Ionization Anomaly (EIA) and the magnetic equator. In this study, we observe the bright flux calibrator 3C48 for ten hours to characterize and study the low-latitude ionosphere with the upgraded GMRT (uGMRT). We outline the methods used for wideband data reduction and processing to accurately measure differential TEC ($\delta {\rm TEC}$) between antenna pairs, achieving a precision of $<$ mTECU (1~mTECU = $10^{-3}$~TECU) for the central square antennas and approximately mTECU for the arm antennas. The measured $\delta {\rm TEC}$ values are used to estimate the TEC gradient across the GMRT arm antennas. We measure the ionospheric phase structure function and find a power-law slope of \( \beta = 1.72 \pm 0.07 \), indicating deviations from pure Kolmogorov turbulence. The inferred diffractive scale, the spatial separation over which the phase variance reaches \(1~\text{rad}^2\), is  $\sim 6.66$~km. Small diffractive scale implies high phase variability across the field of view and reduced temporal coherence, which poses challenges for calibration and imaging.}


\usepackage{etoolbox}
\AtBeginEnvironment{abstract}{\nolinenumbers}

\begin{document}

\nolinenumbers

\section{Introduction}
The ionosphere remains a persistent barrier to radio frequency observations from the ground, with its impact being especially significant in high-resolution interferometric imaging. This highly variable, partially ionized plasma medium introduces several propagation effects, such as refraction, dispersion, group delay, and Faraday rotation \citep{Thompson01,Mangum15}. Major interferometers like the Giant Metrewave Radio Telescope (GMRT)\footnote{\url{http://www.gmrt.ncra.tifr.res.in/}}, the Very Large Array (VLA)\footnote{\url{https://public.nrao.edu/telescopes/vla/}}, the Low-Frequency Array (LOFAR)\footnote{\url{https://www.lofar.org/}}, and the Murchison Widefield Array (MWA)\footnote{\url{http://www.mwatelescope.org/}}, as well as upcoming Square Kilometre Array (SKA) \citep{Fallows20}, are all susceptible to ionospheric disturbances. Although advances in low-frequency radio instrumentation now allow us to investigate ionospheric properties with unprecedented sensitivity and resolution, compensating for ionospheric-induced errors in observational data remains a persistent challenge, particularly for long baselines and low frequencies below 1 GHz.

The ionosphere's behaviour varies spatially and temporally, influenced by solar radiation, geomagnetic activity which is maximum around the geomagnetic equator \citep{OPIO20151640}, and atmospheric conditions. During daytime, enhanced ionization leads to stronger propagation effects, whereas at night the ionosphere becomes more stable; however, it may still add phase errors \citep{Thompson01} to the incoming radio waves. Rapid fluctuations in the total electron content (TEC) can cause signal deflection, scintillation, and differential delays between antennas in an array, which severely affect the coherence of astronomical signals \citep{Lonsdale2005, Intema09}. For precision imaging, especially at frequencies below a few hundred megahertz, accurate modelling and correction of these ionospheric perturbations are essential.

Interferometric arrays such as the VLA (located at approximately 34$^\circ$\,N) have been employed to investigate small-scale ionospheric fluctuations on the order of 0.1 milli Total Electron Content Unit ($1 \, {\rm mTECU} = 10^{-3} \, {\rm TECU}$) by observing a single bright source within the field of view (FoV) \citep{Helmboldt12, Jacobson92}. These studies analyse the ionosphere's behaviour through Fourier analysis of phase measurements from individual antennas. In contrast, wide-field interferometers, capable of detecting multiple celestial radio sources simultaneously, utilize positional offsets to probe ionospheric structures \citep{Cohen09, Jordan17}. Additionally, \citet{Helmboldt_Lane12} identified groups of wave-like disturbances by performing spectral analysis on the position shifts of 29 bright sources within a single FoV observed by the VLA. Using LOFAR (located at mid-latitude), \citet{Mevius16} found that ionospheric irregularities ($\delta$TEC) were anisotropic at night with an accuracy of $1 \times 10^{-3}$ total electron content (TEC).  

The ionosphere behaves as a cold, collision-less plasma, and its refractive index $n$ varies with both the position and time of the observation, making the problem more complex. At frequencies well above the ionosphere's plasma frequency ($\sim$ $1 - 10$ MHz), the refractive index can be expanded using a Taylor series approximation (see, e.g., \citep{Appleton1946,Thompson01}. In this regime, the dominant term in the expansion corresponds to a dispersive phase delay proportional to the total electron content (TEC) along the line of sight. The resulting ionospheric phase shift can be expressed as

\begin{equation}
\phi_{\text{ion}} = \frac{e^2}{4\pi \epsilon_0 c m_e \nu} \int n_e(s) \, ds,
\end{equation}

where $n_e(s)$ is the electron density along the path, $e$ is the elementary charge or electronic charge, $m_e$ is the electron mass, $c$ is the speed of light, and $\epsilon_0$ is the permittivity of free space. The Total Electron Content (TEC) is defined as the integral of the electron number density \( n_e(s) \) along the path \( s \) of a signal through the ionosphere:
\[
\text{TEC} = \int_{\text{path}} n_e(s) \, ds
\]
where \( n_e(s) \) is measured in electrons/m\(^3\), and the TEC is typically expressed in TEC units (TECU), with $\text{TECU} = 10^{16} \, \text{electrons/m}^2$.

Neglecting higher-order terms in the Taylor series of the refractive index \citep{Appleton1946}, the first-order additional phase shift due to the ionosphere can be conveniently expressed as, 

\begin{equation}
\Delta \phi_{\text{ion}} \approx 8.45 \left( \frac{1\,\text{GHz}}{\nu} \right) \left( \frac{\delta \text{TEC}}{1\,\text{TECU}} \right) \quad \text{radians}
\end{equation}

where $\nu$ is the observing frequency in Hz, and TEC refers to the line of sight integrated electron density in units of TECU.

The study of the Earth's ionosphere using radio interferometers has traditionally been constrained by the geographic locations of the telescopes, limiting observations to the local ionospheric regions directly above the arrays. In this context, the Giant Metrewave Radio Telescope (GMRT) offers a distinct advantage because of its favourable geographic location. Located at a latitude of 19$^\circ$05'35.2''\,N and longitude of 74$^\circ$03'01.7''\,E, the GMRT lies in a geophysically sensitive region between the magnetic equator and the northern crest of the Equatorial Ionization Anomaly (EIA), an important feature of the low-latitude ionosphere \citep{Appleton1946}. Ionospheric activity in the EIA regions can also disrupt the operation of the Global Positioning System (GPS). The activation of the Equatorial Ionization Anomaly (EIA) in the evening is mainly caused by a process called pre-reversal enhancement. This happens around sunset, when changes in the electric field and ionospheric dynamics strengthen the upward movement of plasma near the equator. As a result, the two bands (or "crests") of high electron density on either side of the magnetic equator become more pronounced. This evening increase in EIA strength is a key factor in the formation of plasma bubbles and other ionospheric disturbances \citep{Balan18}. \citet{Mangla22} recently used the GMRT to observe the quasar 3C68.2 during a quiet period of solar activity. They found stronger ionospheric fluctuations than those seen with other telescopes like the VLA, LOFAR, and MWA. This is expected because the GMRT is located near the magnetic equator, where the ionosphere is usually more active.

In this study, we focus on closely examining the ionospheric information captured during a single ten-hour observation using the upgraded Giant Metrewave Radio Telescope (uGMRT, \citet{Gupta17})), which is an enhanced version of the original GMRT with improved receivers, wider frequency coverage, and advanced backend systems. The entire observation was done using the Band-4 of uGMRT which operates between 550-750 MHz frequency range.
The observation targets one of the brightest flux calibrators in the sky, 3C48. 3C48 is a bright extra-galactic quasar commonly used for flux density and bandpass calibration in radio astronomy due to its compactness and high brightness (e.g., \citet{Perley13}). Its steady flux density across radio frequencies makes it a reliable calibrator for ionospheric studies.
After the data were flagged and averaged, they were calibrated using the Common Astronomy Software Application (CASA) software package \citep{CASA22} which is a widely used software suite for processing and analysing radio astronomical data, particularly from interferometric arrays. During the calibration process, we estimate the complex bandpass gains by adjusting them to minimize the difference between the observed visibilities and those predicted by a model. The resulting calibration gains capture all distortions that occur between the source and the antenna, including un-modelled variations in the antenna gain pattern, instrument-related errors, and ionospheric effects \citep{Smirnov11}. This study provides a valuable opportunity to develop and refine techniques for future data processing procedures, enhance our understanding of the physical processes driving the observed behaviour, and mitigate major ionospheric-induced systematic effects in uGMRT data. Although multiple Global Positioning System (GPS) stations located near the Murchison Radio-astronomy Observatory are marginally used to calibrate interferometric observations \citep{Arora16}, there are very few GNSS stations available around the GMRT site. The closest known GNSS station is located in Hyderabad, approximately 500 km from the GMRT, making it challenging to characterise fine-scale ionospheric variability. The work presented here represents one of the first systematic investigations of the Earth's ionosphere conducted using the wide-band receiver of the upgraded GMRT.

The structure of this paper is organized as follows. In Section (\ref{sec:observation}), we describe the uGMRT observations and data processing, including Radio Frequency Interference (RFI) mitigation, calibration steps, and phase corrections. Section (\ref{sec:dtec}) outlines the methodology for extracting ionospheric information including estimates of $\delta \text{TEC}$ and $\delta \text{TEC}$ gradients from the calibrated phases. Finally, in Section (\ref{sec:discussion}), we present our discussions and future outlook of this paper.

\section{Observation and Processing}
\label{sec:observation}
The GMRT (Giant Metrewave Radio Telescope) is one of the most sensitive and fully operational radio telescopes designed for low-frequency observations \citep{Swarup1991}. It comprises 30 parabolic dishes, with a diameter of 45 meters, spread across a 25 km region. This configuration provides a total collecting area of around 30,000 m$^2$ at meter wavelengths, which is crucial for sensitive, high-resolution (a few arc-seconds) imaging.

Of the 30 antennas, 14 are positioned randomly within a central square measuring approximately $1.4 \times 1.4$ km$^2$, while the remaining 16 are distributed along three arms extending about 14 km each in a roughly ‘Y’-shaped formation. This layout allows for the study of ionospheric variations over a broad spectrum of spatial scales, as discussed in \citet{Lonsdale2005}. The GMRT’s location (latitude: $19^\circ 05' 35.2''$ N, longitude: $74^\circ 03' 01.7''$ E) places it in a geophysically significant area between the magnetic equator and the northern crest of the EIA \citep{Appleton1946}, making it an excellent instrument for ionospheric studies. Similar to the Very Large Array (VLA), the GMRT’s ‘Y’-shaped configuration enables observations of ionospheric fluctuations in three distinct directions, further enhancing its capability to explore ionospheric dynamics.

We proposed uGMRT observations of 3C48 at Band-4 for 10 hours (proposal code: $47\_003$ \footnote{\url{https://naps.ncra.tifr.res.in/goa/data/search}}),approximately close to the meridian transit of the source. 3C48 is a bright, compact quasar with a flux density of approximately 5.7 Jy at 4.8 GHz \citep{Pearson85}. The bright quasar 3C48 (J2000: RA 01h37m41s, DEC +33d09m35s) was chosen as it dominates the visibilities and enables reliable ionospheric characterization.

We focus exclusively on nighttime ionospheric conditions. Since 3C48 is only observable at night during winter, observations were conducted in November 2024. The dataset covers a total bandwidth of 200 MHz, ranging from 550 MHz to 750 MHz. The integration time for data recording was 2.68 seconds. The full bandwidth was divided into 8192 channels, with a channel resolution of 24.41 kHz. More details about the observations are summarized in Table~\ref{obs_sum}. For ease of computation, in this study we used data from 3,908 channels, each channel had a bandwidth of 24.4 kHz, spanning frequencies from 553 MHz to 648 MHz, with a central frequency near 600 MHz.

\begin{table}[h]
    \centering
    \caption{Observation summary}
    \label{obs_sum}
    \begin{tabular}{l c}
        \hline
        Project Code & 47\_003 \\
        Observation Date & 23 November 2024 \\
        Start and End Time of Observation (IST) & 5:21 PM to 3:27 AM \\
        Bandwidth & 200 MHz \\
        Central Frequency & 650 MHz \\
        Number of Channels & 8192 \\
        Frequency Resolution & 24.41 kHz \\
        Integration Time & 2.68 seconds \\
        Correlation Products & RR and LL \\
        Number of Working Antennas & 30 \\
        Target Source & 3C48 \\
        RA (J2000) & 01$^h$37$^m$41.301596$^s$ \\
        DEC (J2000) & +33$^\circ$09$'$35.25459$''$ \\
        Total On-Source Time & 10 h \\
        \hline
    \end{tabular}
\end{table}

\subsection{RFI mitigation}

Radio frequency interference (RFI) degrades the sensitivity of radio observations by increasing system noise and distorting calibration solutions. It also restricts the usable frequency bandwidth, with particularly severe effects at sub-GHz frequencies in GMRT data.

To mitigate RFI, we first remove strong interference and then apply an automated flagging algorithm using the TFCROP task from the Common Astronomy Software Applications (CASA) package \citep{CASA22}. This algorithm identifies outliers in the 2D time-frequency space for each baseline. It operates iteratively, using a user-defined sliding time window to establish a clean band-shape template that excludes RFI-contaminated regions.

During each iteration, the algorithm calculates the standard deviation between the data and the fitted template, flagging points that exceed a set threshold: 4-sigma in the time domain and 3-sigma in the frequency domain, which is optimized for strong narrow-band RFI. Additionally, we enable the EXTEND mode in CASA to flag data when TFCROP identifies RFI in at least $50 \, \%$ of the time-frequency bins. Around $30 \, \%$ of data was flagged for both  Right-hand (RR) and Left-hand (LL) circular polarizations after TFCROP task, rest is some iterative flagging based on the calibration solutions.

\subsection{Calibration steps}

After identifying and removing spurious signals from the dataset, we perform direction-independent calibration using standard CASA tasks. The source 3C48 is used as the flux density, bandpass, and phase calibrator. Note that 3C48 is also our target source. To set its flux density, we rely on the \citet{Scaife12} model, applying it using the CASA task SETJY,  which set the flux density (in Jansky units) of a calibrator source.

The calibration process begins with delay calibration (CASA task GAINCAL with gaintype=`K'), where delay corrections are calculated relative to a reference antenna (here we used a central square antenna `C06'), which has its phase assumed to be zero. This removes any linear phase variation with frequency. Next, we perform bandpass calibration with solint = `8ch, 10s' using the CASA task BANDPASS. This solves for the antenna gain solutions with a time-frequency solution interval of 10s and $\sim 194$ kHz respectively. In general, the bandpass corrections are complex frequency dependent phase and amplitude corrections (here time dependent also), for the remainder of our analysis, we only concentrate on the phase solutions. The derived  bandpass phase solutions will capture both the instrumental response and ionospheric effects. If the ionospheric variations are more significant we can use shorter time interval and narrower channels with higher frequency resolution \citep{Gasperin18}.

The selection of a solution interval can be tricky, but it is all about the required frequency or timescales. On an integration-to-integration (a few seconds)per channel basis, our visibilities will be dominated by thermal noise, while on longer timescales or frequency resolutions, we can average more data so that instrumental and line-of-sight effects can be averaged out, and we may not pick up the ionospheric trends accurately. Here, we set our frequency and time scales to achieve a good signal-to-noise ratio (SNR, at least 3 per baseline) and also made sure that the bandpass solutions converge. We also expect the instrument to be relatively stable at the time-frequency resolution of the estimated gain solutions. During this process, antennas with large errors in complex gain solutions are flagged, along with specific baselines for certain scans. After applying calibration and RFI mitigation to the target field, approximately $32 \%$ of the data is flagged.

\subsection{Phase corrections}
To quantify the phase variation of a radio signal as it passes through the Earth's ionosphere, we use the method described by \citet{Helmboldt12}. The phase values are initially estimated using CASA, followed by a series of steps applied to the phase data to extract ionospheric information.
In general, the phase measurements also include contributions from other sources, but the ionospheric effect is more dominant at lower radio frequencies.

The difference in phase between two antenna elements, as explained by \citep[see][]{Helmboldt12, Mangla22}, is expressed as:

\begin{eqnarray}
        \Delta\phi = \Delta\phi_{\rm ion} + \Delta\phi_{\rm instr} + \Delta\phi_{\rm amb} + \Delta\phi_{\rm sour} 
    \label{eq:phase_diff}
\end{eqnarray}
where $\Delta\phi_{\rm ion}$ is the difference in the ionospheric phases of the two antennas along the line of sight, $\Delta\phi_{\rm instr}$ represent the difference in the instrumental effects between the two antennas, $\Delta\phi_{\rm amb}$ is the contribution from $2\uppi$ ambiguities and $\Delta\phi_{\rm sour}$ is the phase contribution due to the structure of the observed source. One can derive a model of the target source from higher resolution observations (such as clean component model or Gaussian source model). The the model visibilities can be computed and $\Delta\phi_{\rm sour} = arg(V_{model})$, where $arg(V_{model})$ is the phase of the complex model visibility. Once, we have $V_{model}$ we can divide the observed visibility by the model to remove the source phase contribution. $\Delta\phi_{\rm amb}$ can be removed by having a short time sampling to ``unwrap'' the phases. 

\begin{figure}
    \centering
    \includegraphics[width=\linewidth]{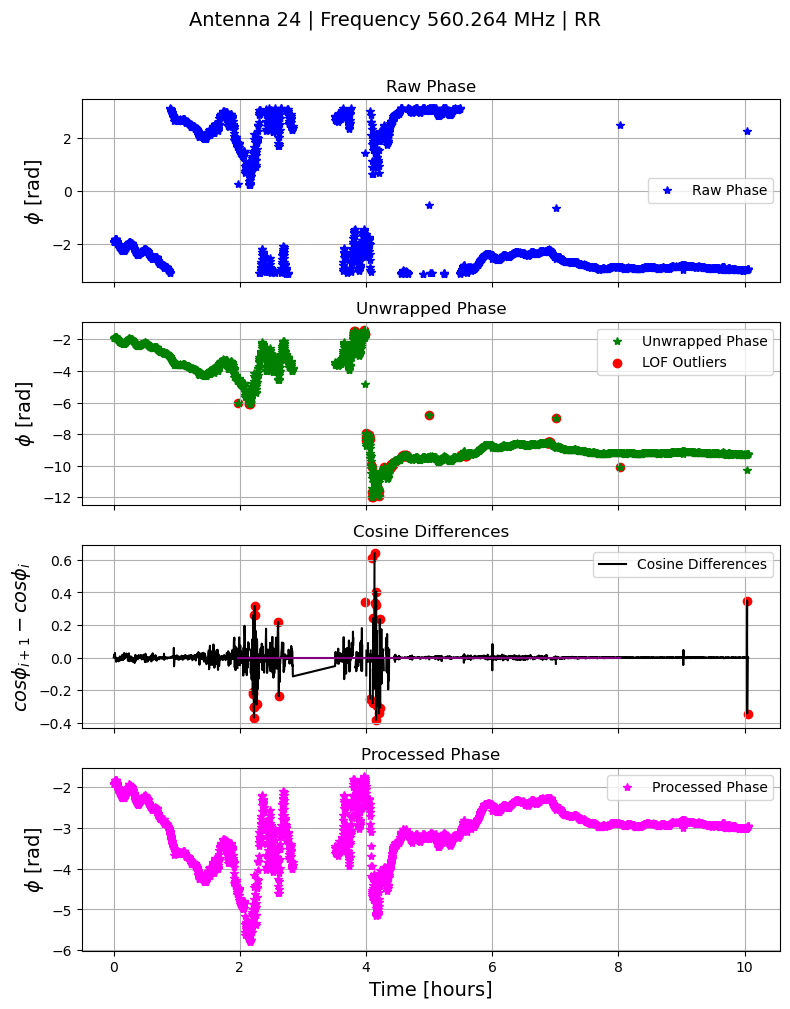}
    \caption{(1st row) The raw phase at 560.26 MHz for antenna 24, RR polarization, as a function of time. 
    (2nd row) The unwrapped phases with LOF outliers marked with red circles.
    (3rd row) The difference between the cosine of the wrapped phase at a time step $i$ and 
    the next time step $i+1$ as a function of time, used to find spikes in the phase data. 
    Flagged spikes are highlighted with red circles. (4th row) The processed unwrapped phases where outliers are flagged at $5 \, \sigma$ level.}
    \label{fig:cosLOFflag}
\end{figure}

To remove outliers and phase jumps in time (Figure~\ref{fig:cosLOFflag}), we used a Local Outlier Factor (LOF) method and a cosine filter in time  to detect and remove anomalous time jumps in phase \citep{Helmboldt12, Mangla22}. The Local Outlier Factor (LOF) algorithm measures the local deviation of a data point with respect to its neighbours. It identifies outliers based on how isolated they are from their surroundings.

The LOF score for time step t is defined is defined as, \[
LOF_k(t) = \frac{ \sum_{q \in N_k(t)} \frac{ l_k(q) }{ l_k(t) } }{|N_k(t)|}
\]
where \( N_k(t) \) represents the set of \( k \)-nearest neighbours in time, and \( l_k(t) \) is the local reachability density given by:\[
l_k(t) = \frac{|N_k(t)|}{\sum_{q \in N_k(t)} d_k(t,q)}
\]

where \( d_k(t,q) \) is the Euclidean distance in phase space. We choose $\rm{N_k(t)} = 100$. We use the function {\it LocalOutlierFactor(\texttt{n\_neighbors=100, contamination="auto"})} from scikit-learn \footnote{\url{https://scikit-learn.org/stable/}}. Time steps with anomalous LOF scores were flagged and removed, and the missing data points were linearly interpolated to ensure temporal continuity.

Following \citet{Helmboldt12} to remove short phase jumps from the unwrapped phases (Figure~\ref{fig:cosLOFflag}, 2nd row), we designed a cosine filter to remove spurious jumps in the wrapped phase before unwrapping it. The idea is to track how much $cos(\phi)$ — where $\phi$ is the wrapped phase —changes from one time step to the next. If the change at a particular time step is unusually large (specifically, exceeding five times the standard deviation across all time steps for a given antenna), we flag that point as unreliable. These flagged time steps are left out during the phase unwrapping. Later, any gaps caused by the flagged points are filled in by interpolating the unwrapped phase values from the surrounding good data points. The final, cleaned phase solutions exhibit a substantial reduction in phase jumps and outliers. 
An example of this improvement for antenna 24 is shown in the bottom row of Figure~\ref{fig:cosLOFflag}.

\begin{figure}[H]
    \centering
    \includegraphics[width=\linewidth]{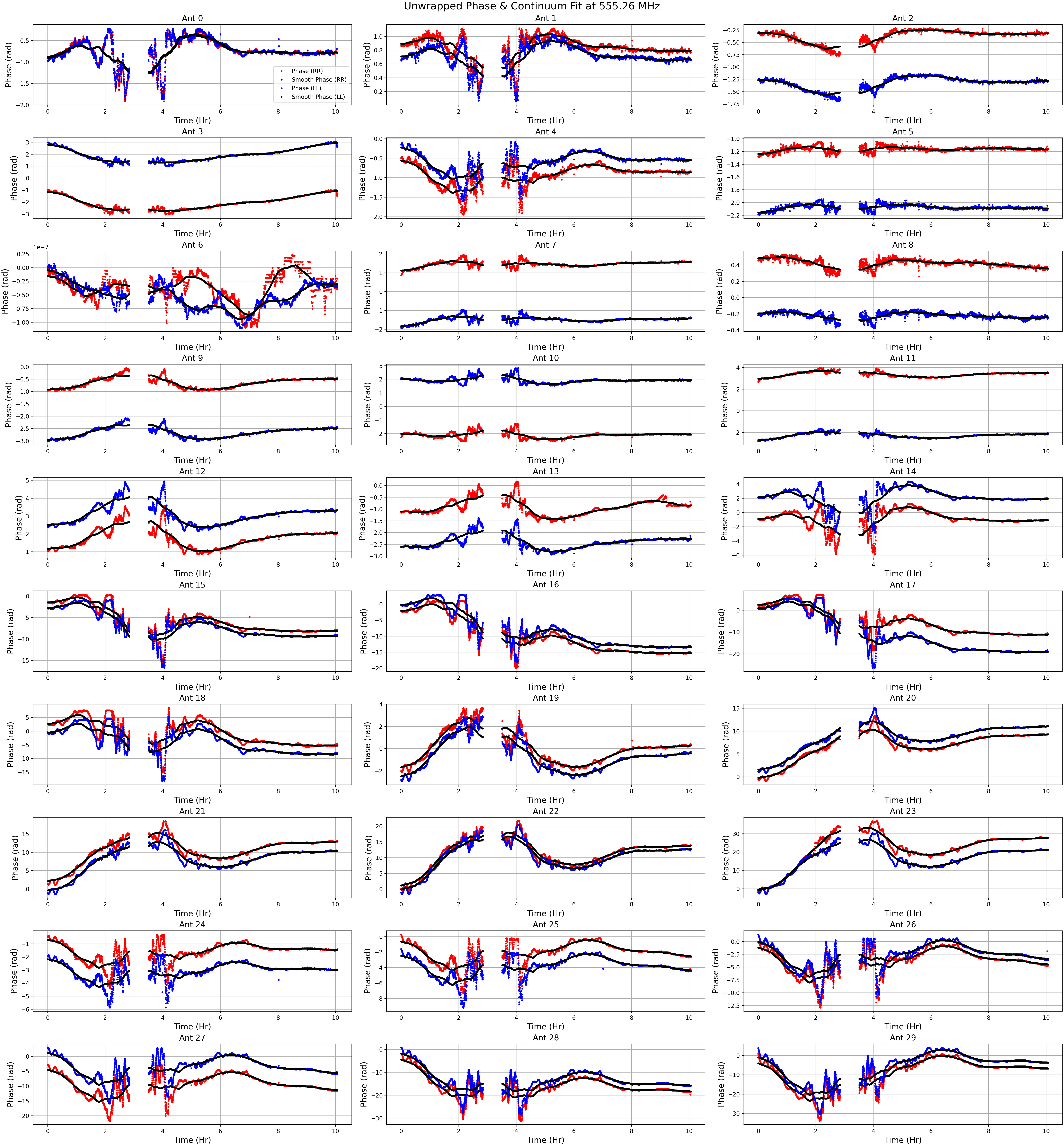}
    \caption{This figure shows the unwrapped phases for polarization RR and LL, smoothed using a one-hour-wide boxcar window, which effectively preserves apparent fluctuations while providing a clear representation of the continuum. 
    The red and blue points represent the unwrapped phases for correlation RR and LL respectively, while the solid black lines show the smooth fits at 555.26 MHz.}
    \label{fig:smoothfit}
\end{figure}

The instrumental phase includes the error associated with the delays, the actual source position and the offset between antenna line of sight. These vary relatively slowly in time comparatively to the ionospheric time fluctuations. We model instrumental effects using a continuum subtraction approach following \citet{Helmboldt12}. We treated ionospheric fluctuations as features superimposed on a smooth continuum consisting of the instrumental phases, which vary relatively slowly with time. 
We performed continuum subtraction for each antenna, band, and polarization by smoothing the unwrapped phases using a one-hour-wide boxcar window (Figure~\ref{fig:smoothfit}). This method effectively preserved apparent fluctuations while providing a reliable representation of the continuum. Unfortunately, we lack the capability to separate the slowly varying component of the ionospheric phase from instrumental effects. As a result, we can only measure fluctuations in TEC on relatively short timescales ($\le$ 1 hour). 

\begin{figure}
    \centering
    \includegraphics[width=\linewidth]{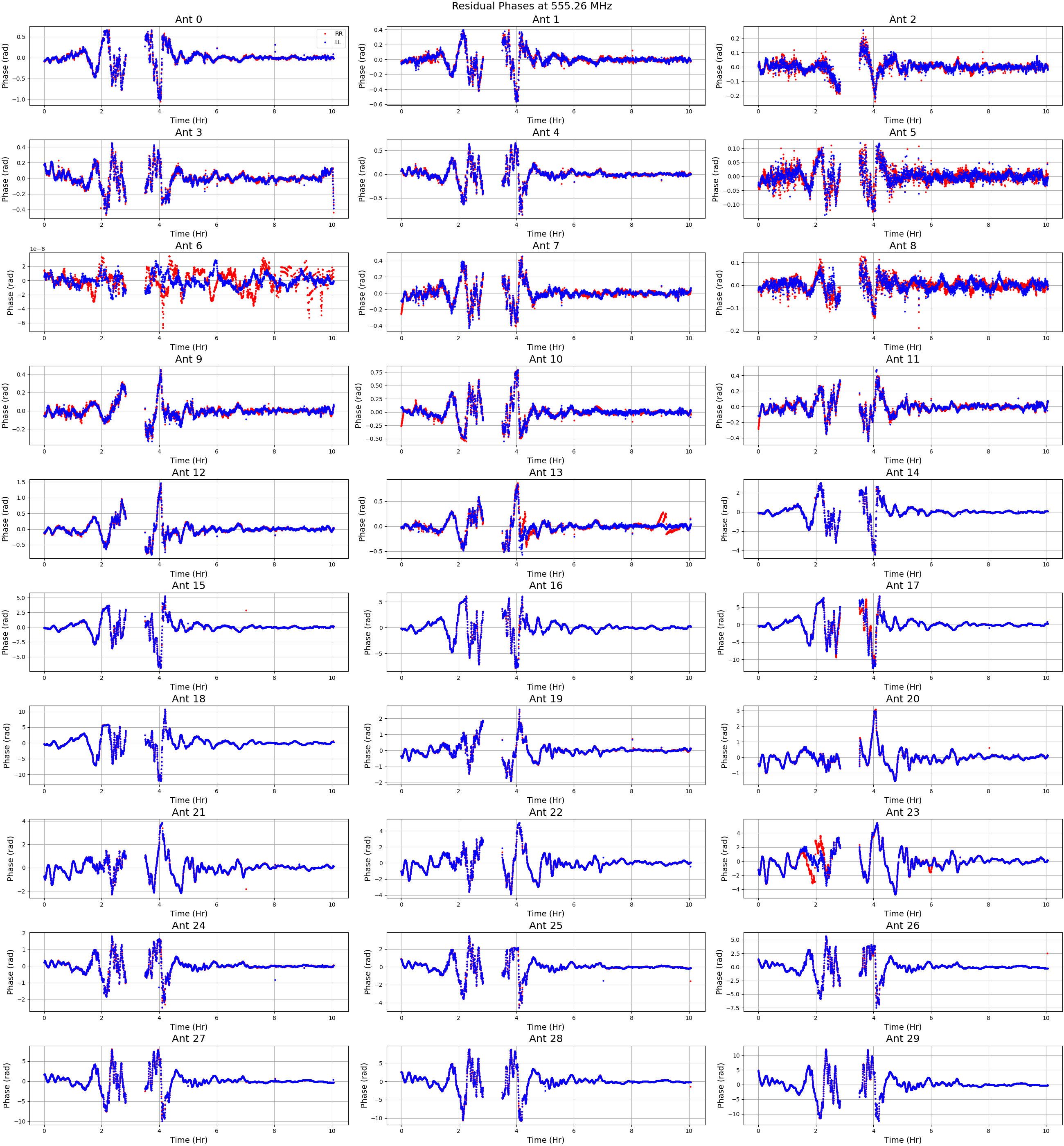}
    \caption{The residual phases after the smooth continuum is removed at 555.26 MHz.}
    \label{fig:residual}
\end{figure}

Furthermore, to estimate the smooth continuum component of the instrumental phase, we can utilize the two ends of the wide-band data — narrow frequency ranges ($\sim$ 10 MHz each) separated by several MHz. The phase difference between the lower and upper sub-bands ($\phi_{\rm low}$ and $\phi_{\rm up}$), is computed and scaled by $\frac{\nu_{\rm up}}{\nu_{\rm low}}$, effectively suppressing ionospheric contributions that scale as 
$\nu^{-1}$. 
This procedure isolates the instrumental phase difference, as described in \citet{Mangla22}.
However, for our current wide-band dataset, we used the full frequency range and choose to fit a smooth model over time for each frequency channel. This method allows us to better account for how the instrument's response changes with frequency and to capture a clean, smooth representation of the continuum signal over time \citep{Helmboldt12}. We find the residual phases after continuum subtraction varies randomly around mean 0, with a scatter of $\pm 0.5$ radians for the central square GMRT antennas, whereas for the arm antennas the maximum scatter in residual phases is around $\pm 5.0$ radians (Figure~\ref{fig:residual}). 
We encountered a period of rapid phase and amplitude variation between approximately 19:30 and 21:45 IST. This significantly degraded the data quality, particularly during the interval from 20:15 to 20:55 IST (IST), which corresponds to the $\sim$ 45-minute gap in our analysis. The possible causes are radio frequency interference (RFI) or some instrumental issues (drift due to LO (local oscillator) unlock, bank-end issues etc.) which can also cause rapid phase jumps. Also, these fluctuations were significantly stronger than the neighbouring time steps, so we considered the data to be unreliable and decided to flag this time range.
Residual RFIs may have impacted nearby time ranges, and to a large extent, the residual phases show increased scatter for all antennas within the 2 to 4 hour range compared to the rest of the night's data. It is also possible that the ionosphere was more active during this period, leading to the detection of large-scale traveling ionospheric disturbances (TIDs) or corruption of the visibility phases by ionospheric scintillations \citep{Gasperin18, Fallows20}. Further, around evening time in the Indian sector, both EIA crests intensify due to sunset conditions. This enhancement leads to the development of plasma bubbles, which can spread across the EIA trough and the collapsing crests \citep{Balan18}. Strong scintillations during this time were also reported in the GMRT observation log file.

\section{Obtaining Ionospheric Information From Calibration Phases}
\label{sec:dtec}

After removing the instrumental component, we convert the residual phases for each antenna and polarization to differential TEC ($\delta {\rm TEC}$) using the scaling factor $\nu/(8.45 \times 10^9)$, where $\nu$ is in units of Hz. We calculate the median $\delta {\rm TEC}$ using all available values across frequency channels and both polarizations. To estimate the uncertainty, we use the median absolute deviation (MAD), computed at each time step for each antenna across all available frequency channels and polarization data (898 data points per time step). We report the median $\delta {\rm TEC}$ values to reduce the influence of any remaining outliers in the data.

\begin{figure}
    \centering
    \includegraphics[width=\linewidth]{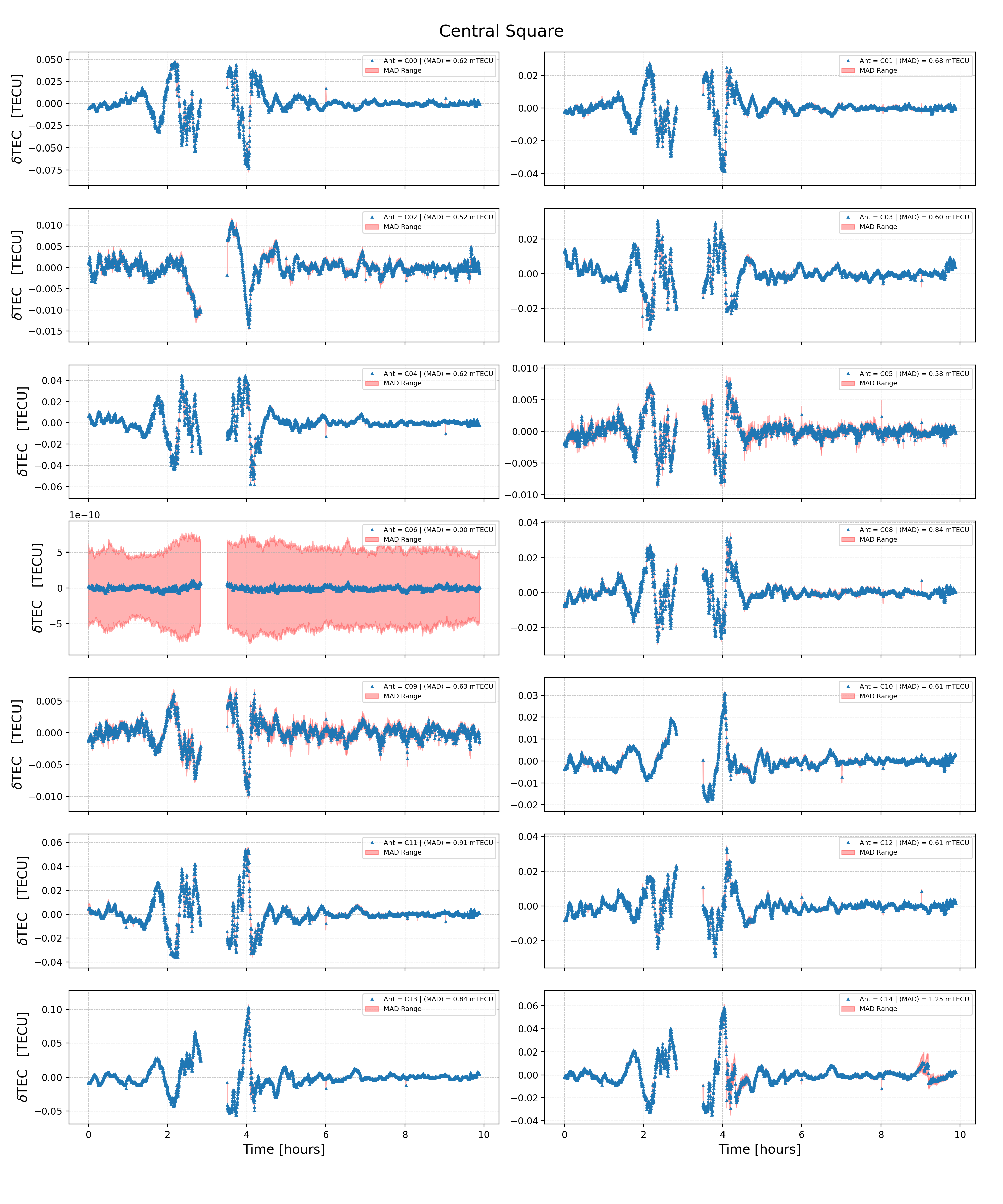}
    \caption{For each antenna in the GMRT central square, this figure shows the $\delta {\rm TEC}$ along the antenna's line of sight with respect to that of the reference antenna (`C06'). The MAD values are also displayed in each panel to represent the estimated uncertainty.}
    \label{fig:dtec_antcs}
\end{figure}

\begin{figure}
    \centering
    \includegraphics[width=\linewidth]{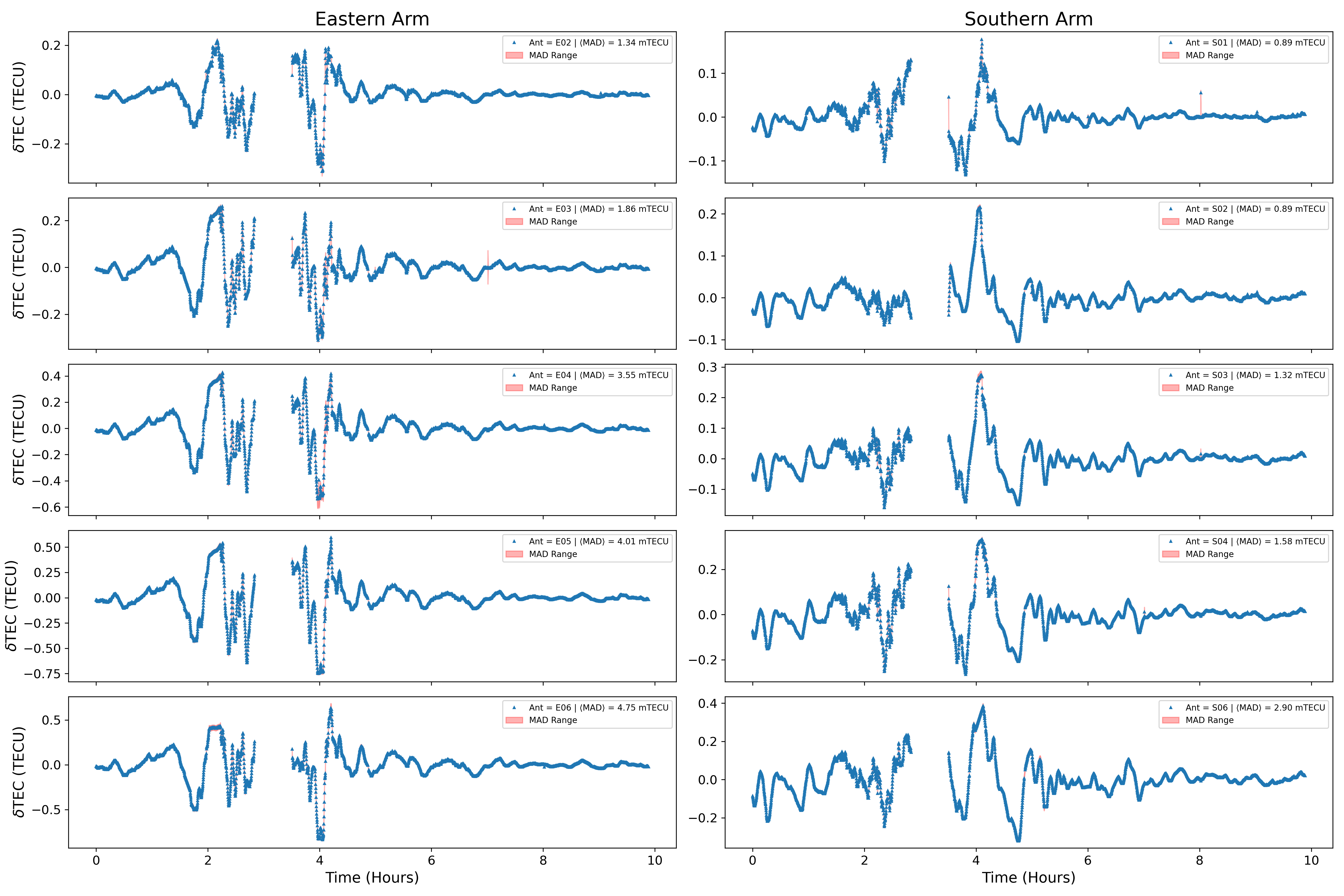}
    \caption{Similar to \ref{fig:dtec_antcs}, but along the eastern and southern arm of GMRT.}
    \label{fig:dtec_antestsouth}
\end{figure}

\begin{figure}
    \centering
    \includegraphics[width=\linewidth]{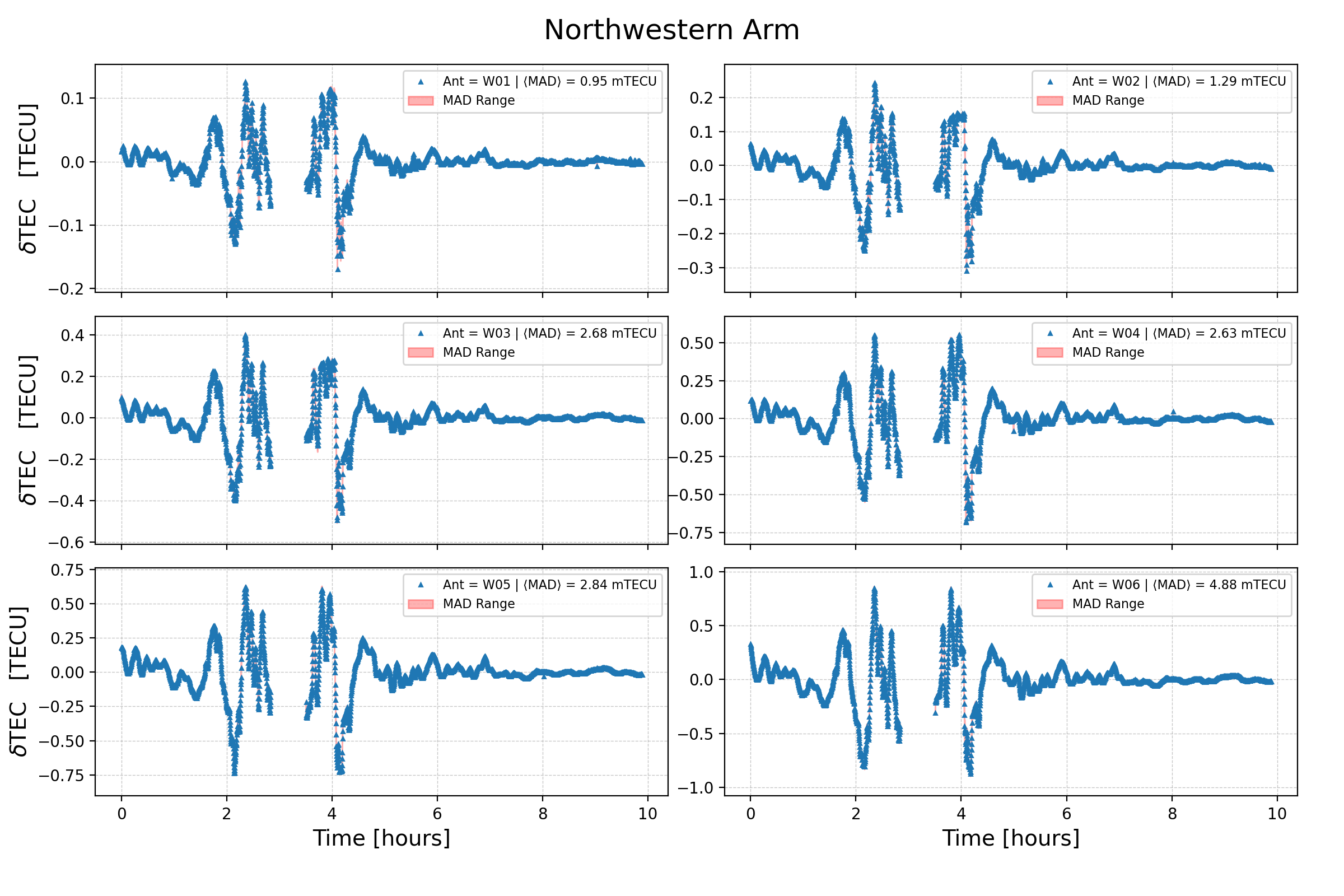}
    \caption{Analogous to \ref{fig:dtec_antcs}, but now along the northwestern arm of GMRT.}
    \label{fig:dtec_antnorth}
\end{figure}

The resulting $\delta {\rm TEC}$ values for each antenna as a function of time are shown in Figure~\ref{fig:dtec_antcs} for the central square antennas, in Figure~\ref{fig:dtec_antestsouth} for the eastern and southern antennas, and in Figure \ref{fig:dtec_antnorth} for the northwestern arm. The uGMRT data typically achieve less than mTECU ($\leq$ mTECU) uncertainty for the central square antennas, whereas the variation, represented by the MAD values, is around a few mTECU ((1~mTECU = $10^{-3}$~TECU)) for the arm antennas.

To independently verify whether our $\delta {\rm TEC}$ estimates are affected by residual RFI, we analysed a subset of data spanning the frequency range 590–615 MHz using AOFlagger \citep{Offringa12} — instead of the CASA flagging routine — and applied a Lua strategy file\footnote{\url{https://drive.google.com/file/d/1zrqessbSLLrn8mpIhpQlOw4AjN4EIlkm/view?usp=drivesdk}} for RFI detection and removal. We observed very similar $\delta {\rm TEC}$ variations over time for both the central square (Figure~\ref{fig:tec_censq_casa_aoflag}) and arm antennas of the GMRT. Based on this consistency, we proceeded with the CASA-flagged data for the remainder of our analysis.

The differential TEC measured along the three arms of the GMRT exhibits slightly different patterns. In general, the arm antennas show stronger and slower fluctuations, which are characteristic features of travelling ionospheric disturbances (TIDs) \citep{Her06}. This behaviour could result from a large-scale wave moving through the ionosphere along with smaller waves, or from multiple waves propagating in different directions. Later, when we compute the spatial gradient (Section \ref{sec:tec_grad}), we aim to better decipher the dominant component—whether east-west or north-south—of the TIDs. We plan to explore these patterns further using spectral analysis in future work \citep{Helmboldt12b, Mangla23}.

Unwrapped, continuum-subtracted phase differences between antennas can be used to track the temporal variation of $\delta {\rm TEC}$ across the uGMRT baselines. We fit the phase difference using the model $\delta\phi_{ij}(\nu) = - C_1 \, \delta TEC_{ij}/\nu \, \, \mathrm{rad}$, Here, \( C_1 \approx 8.45 \times 10^{9} \, \mathrm{m^{2}s^{-1}} \) represents the ionospheric conversion factor, \( \nu \) is the observing frequency in Hz, and \( \delta TEC_{ij} \) denotes the difference in integrated ionospheric electron content (in TECU) along the lines of sight to antennas \( i \) and \( j \) \citep{Mevius16}. For instruments with large bandwidths, this model allows us to exploit the wavelength dependence to better separate residual instrumental phases (which scale as $\propto \nu$) from the ionospheric phase (which scales as $\propto \nu^{-1}$).

Two essential geometric corrections must be applied to the antenna positions and $\delta TEC_{ij}$ measurements to more accurately reflect the actual ionospheric conditions. Since the observations of 3C48 include times when the source is relatively close to the horizon, a plane-parallel approximation is insufficient. Instead, a thin-shell approximation is used, in which the ionosphere is modelled as a thin layer at the height of maximum electron density, $z_{\rm ion}$, as determined by the International Reference Ionosphere (IRI) model \footnote{\url {https://www.ionolab.org/iriplasonline/}} \citep{Bilitza22} for the corresponding dates and times of the observations (see Figure~\ref{fig:ion_height_cosep}).
 
The first correction involves converting the ground-based antenna positions to their projected positions within the ionosphere. In a non-plane-parallel atmosphere, these positions vary with the elevation of the observed source. For a spherical ionospheric shell, a "pierce point" can be defined for each antenna, where the line of sight to the source intersects the ionosphere. The relative positions of these pierce points, referenced to the centre of the array, serve as the projected positions within the ionosphere. We follow Appendix A of \citet{Helmboldt12} to estimate the projected GMRT antenna coordinates onto the ionospheric thin shell at each time step during the observation.

The second correction accounts for the increased path length through the ionosphere when the observed source is closer to the horizon. We converted the slant TEC (sTEC) values to vertical TEC (vTEC) by applying a slant factor: ${\rm vTEC} = \cos(\epsilon) \, {\rm sTEC}$, where $\epsilon$ is the angle between the line of sight from the GMRT to the source and the vertical line from the ionospheric pierce point to the point on the Earth's surface directly beneath it \citep{Helmboldt12}.

\begin{figure}[H]
    \centering    
    \includegraphics[width=\linewidth, height=0.25\textheight]{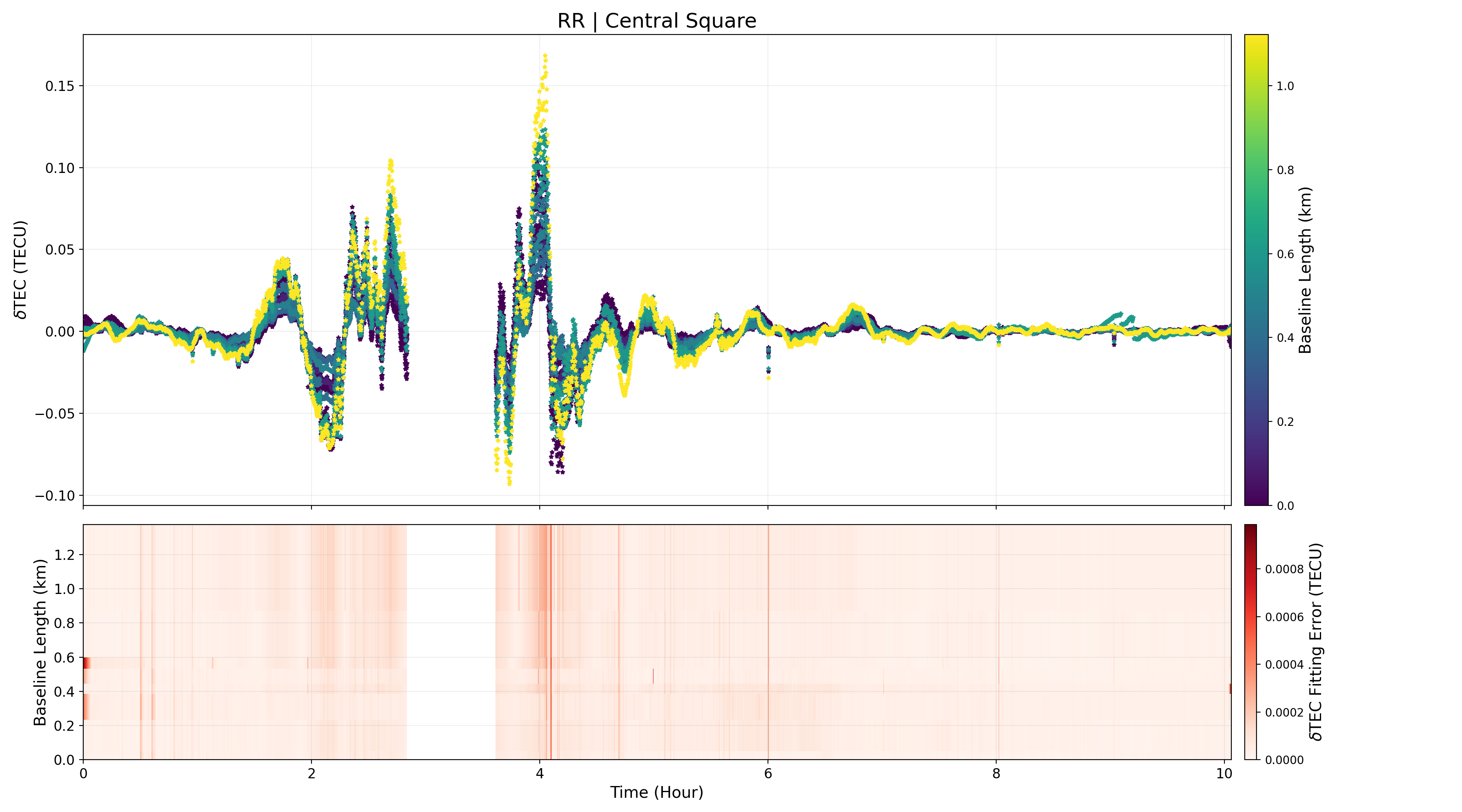}
    \includegraphics[width=\linewidth, height=0.25\textheight]{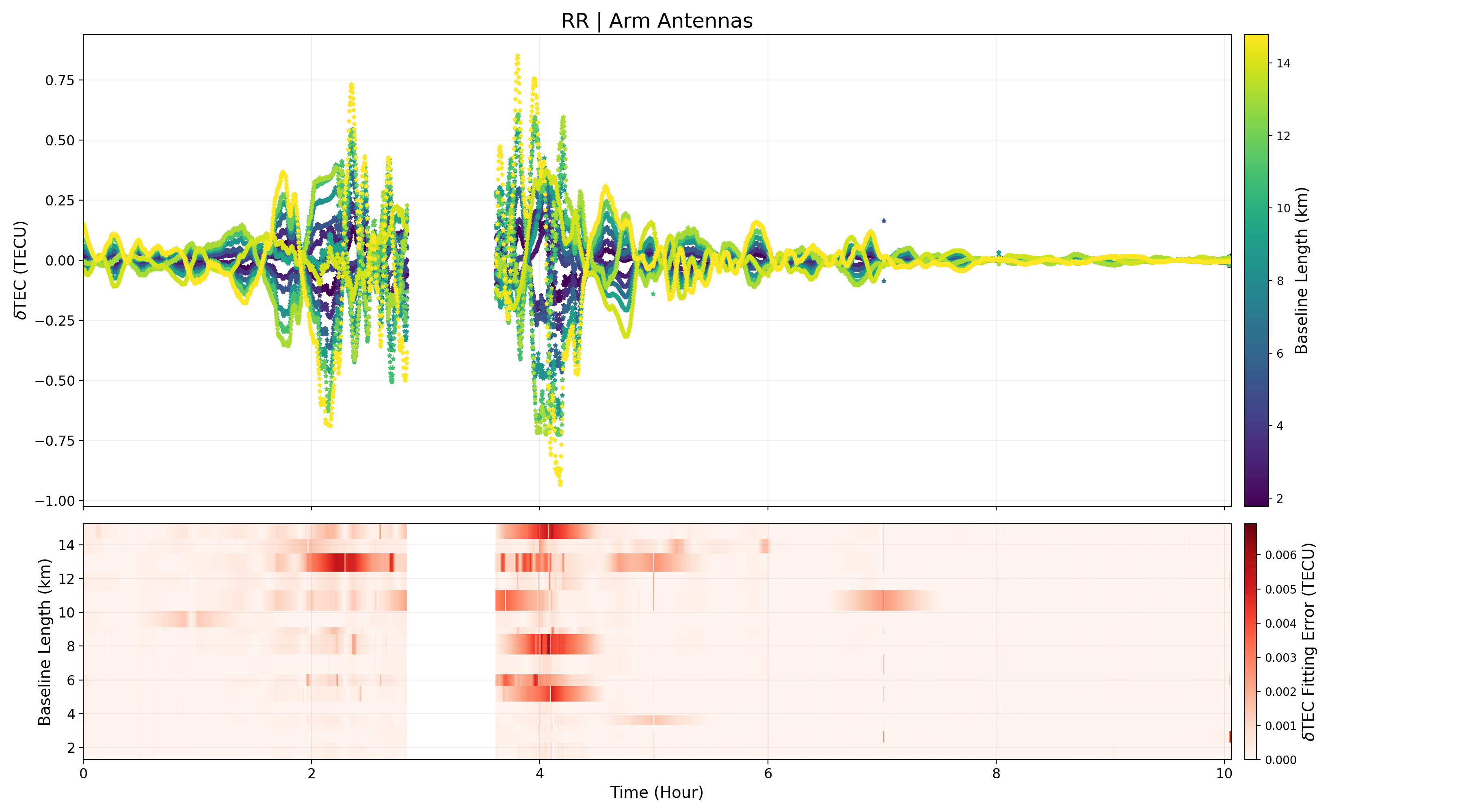}    
    \caption{Differential TEC ($\delta {\rm TEC}$) of uGMRT antennas relative to the refant antenna `C06' over time. The colour-bar represents baseline length. (Top row) Central square baselines. (Bottom row) Longer arm baselines.}
    \label{fig:dtec_base}
\end{figure}

Figure~\ref{fig:dtec_base} shows a representative example of the fitted differential vTEC values for both the central square antennas (baselines $\leq$ 1.4 km) and the arm antennas, using antenna `C06' as the reference. The colour scale represents baseline length. The fitting errors are displayed in the bottom panel of Figure~\ref{fig:dtec_base}. We observe temporally varying ionospheric effects that are spatially coherent across antennas, even on short baselines (~1 km). A Principal Component Analysis (PCA) to the time series of differential TEC ($\delta \rm{TEC}$) showed that just five principal components explained nearly $100 \%$ of the total variance. Typically, for the core baselines, the $\delta {\rm TEC}$ values vary within $\pm$0.1 TECU, whereas for the remote baselines (baselines $\geq$ 1.4 km), the variation is approximately $\pm$1.0 TECU. The fitting errors in $\delta {\rm TEC}$ are about 100 times smaller (see Figure~\ref{fig:dtec_base}). For the central square baselines, the uncertainty (MAD) is around 0.0023 TECU, increasing to $\sim$0.0147 TECU for the arm antennas.

It is important to note that the accuracy of the TEC solutions derived using this method is limited by the second-order phase effects \citep{Mevius16}, including non-linear phase behaviour with frequency, imperfections in the sky model, and cable reflections. Such higher-order effects tend to be more significant at frequencies below 100 MHz \citep{Gasperin18}.

Results from the MWA interferometer have shown that accurately modeling small-scale ionospheric features requires a significantly denser network of Global Navigation Satellite System (GNSS) stations. However, the region surrounding the GMRT site has very few such stations, the closest known GNSS station with publicly available data is located in Hyderabad, approximately 500 km away. This sparse coverage makes it challenging to capture fine-scale ionospheric variability in the area.

Although Total Electron Content (TEC) maps provided by the International GNSS Service \citep{Orus05, Hern09} can be used to infer absolute TEC values (Figure \ref{fig:vtec_igs}), their effectiveness is limited. This limitation arises due to uncertainties in the projected magnetic field parameters and the relatively low temporal resolution of these models, which ranges from tens of minutes to hours. As a result, they are not ideal for assessing real-time ionospheric conditions during low-frequency interferometric radio observations.

Moreover, at Band-4 frequencies, our datasets lack the sensitivity needed to estimate second-order ionospheric effects, such as those caused by Faraday rotation. This further complicates efforts to derive absolute vTEC values from differential TEC ($\delta \rm{TEC}$) measurements \citep{Gasperin18}.

\subsection{TEC Gradients}
\label{sec:tec_grad}
The GMRT measures differential TEC between pairs of antennas. As a result, it is inherently sensitive only to variations in the TEC gradient. However, the geometry of the array makes it difficult to determine the absolute TEC gradient at individual antenna locations.

To overcome this, reconstructing the full TEC gradient requires a somewhat heuristic approach \citep{Helmboldt12, Mangla22}. This reconstruction step is essential for accurately characterizing TEC fluctuations. Without it, the raw $\delta \mathrm{TEC}$ time series can only be interpreted meaningfully in the spectral domain. Even then, strong assumptions — such as the presence of a single plane wave — must be made \citep{Jacobson92}.

Unlike standard interferometric imaging—which involves transforming and deconvolving visibility data to reconstruct sky images—studying ionospheric gradients demands accounting for how each antenna line of sight intersects the ionospheric shell. This introduces an additional layer of geometric complexity. Nonetheless, by leveraging the rotation of the Earth and the drift of ionospheric irregularities, we can map temporal fluctuations into spatial structures. However, in contrast to predictable Earth rotation synthesis, the irregular drift speeds and directions of ionospheric features introduce significant mapping uncertainties. A widely adopted strategy involves decomposing the $\delta \rm{TEC}$ time series into spectral components and analysing how these spectral modes manifest spatially across the array. This method facilitates estimation of key parameters of ionospheric features—such as their scale, drift velocity, and propagation angle \citep{Helmboldt12b, Mangla23}.

As outlined in Section~\ref{sec:dtec}, two geometric transformations were employed to bring the observed gradients in line with their vertical equivalents. First, antenna coordinates were projected onto their corresponding ionospheric pierce points. Second, a slant-to-vertical correction was applied to compensate for the varying elevation angle of the source 3C48 during the observation period.

After applying geometric corrections to both the antenna positions and the $\delta \mathrm{TEC}$ values, we aimed to reconstruct the full two-dimensional TEC gradient at each antenna and time step.
We did not assume a specific structure, such as a plane wave. Instead, we took advantage of the fact that the GMRT array is small compared to the typical scale of ionospheric disturbances.
This allowed us to approximate the TEC surface using a low-order Taylor expansion. We examined the data over several time steps. The results showed that the TEC structure is well represented by a second-order, two-dimensional Taylor series. The corresponding polynomial form is:

\begin{equation}
\rm{TEC} = p_{0}x + p_{1}y + p_{2}x^{2} + p_{3}y^{2} + p_{4}xy + p_{5}
\end{equation}

Here, \(x\) and \(y\) represent the antenna positions in the north-south and east-west directions, respectively. These positions are projected onto the ionospheric surface. The projection is done at the estimated peak height of the ionosphere, as determined by the IRI model.

For robust parameter estimation, we used the \(\delta \mathrm{TEC}\) differences from all 406 unique antenna pairs at each time step. This allowed us to fully utilize the available data.

We model the differential TEC between antenna pairs \( (i, j) \) using a second-order polynomial expansion in the projected antenna coordinates:

\begin{eqnarray}
\delta \mathrm{TEC}_i - \delta \mathrm{TEC}_j &=& p_0 (x_i - x_j) + p_1 (y_i - y_j) 
\nonumber \\
& & + p_2 (x_i^2 - x_j^2) + p_3 (y_i^2 - y_j^2) + p_4 (x_i y_i - x_j y_j)
\label{eq:pfit}
\end{eqnarray}

This can be written in matrix form as:

\begin{equation}
\mathbf{A} \mathbf{p} = \mathbf{b}
\end{equation}

where \( \mathbf{p} = [p_0, p_1, p_2, p_3, p_4]^T \) is the polynomial coefficient vector, \( \mathbf{b} \) is the vector of observed differential TEC values, and \( \mathbf{A} \) is the mapping matrix, with rows based on the antenna coordinates as given in Equation~\ref{eq:pfit}.

We also account for the uncertainties in the derived \(\delta \mathrm{TEC}\) values. To optimize the polynomial coefficients over time, we apply a weighted least squares method. The weight matrix \(\mathbf{W}\) is a diagonal matrix. Each diagonal element corresponds to the inverse of the variance of the differential TEC measurements:

\begin{equation}
\mathbf{W} = \text{diag} \left( \frac{1}{\sigma_1^2}, \frac{1}{\sigma_2^2}, \ldots, \frac{1}{\sigma_n^2} \right)
\end{equation}

Here, $\sigma_i$ denotes the standard deviation of the $i$-th differential TEC measurement. It is worth noting that additional residual systematics may contribute to the overall uncertainty; however, we neglect these effects in our analysis, as they are difficult to quantify a priori.

We solve the following regularized least squares problem, using a small regularization parameter $\lambda = 1 \times 10^{-6}$ for numerical stability:

\begin{equation}
\min_{\mathbf{p}} \left[ (\mathbf{b} - \mathbf{A} \mathbf{p})^T \mathbf{W} (\mathbf{b} - \mathbf{A} \mathbf{p}) + \lambda \|\mathbf{p}\|^2 \right]
\end{equation}

which yields the solution:

\begin{equation}
\mathbf{p} = (\mathbf{A}^T \mathbf{W} \mathbf{A} + \lambda \mathbf{I})^{-1} \mathbf{A}^T \mathbf{W} \mathbf{b}
\end{equation}

We applied standard sigma-clipping during the fitting process. Antenna pairs with absolute residuals greater than \(3 \times \text{rms}\) were rejected at each time step. The rms of the fit residuals was calculated. This procedure was repeated up to 10 times. Between 0 and a maximum of 63 baselines were rejected per time step. To preserve small temporal and spatial changes, each time step was fitted independently.

\begin{figure}[H]
    \centering
    \includegraphics[width=\linewidth, height=0.45\textheight]{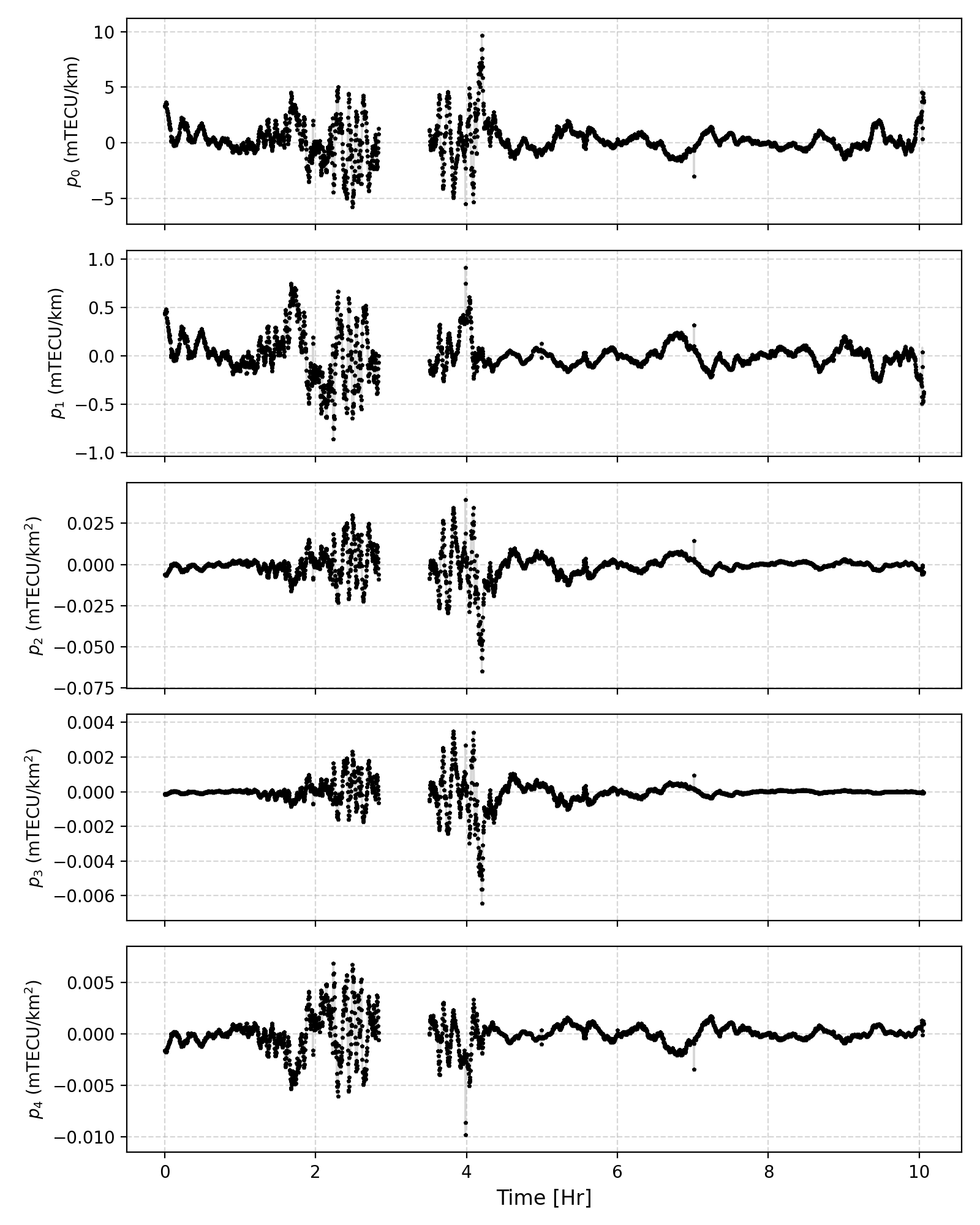}
    \caption{The fitted polynomial coefficients (Eqn. \ref{eq:pfit}) as a function of time.}
    \label{fig:grad_pfit}
\end{figure}

Figure~\ref{fig:grad_pfit} displays the evolution of the fitted coefficients over time. Notably, $p_0$ and $p_1$ — representing the linear TEC gradients in the north-south and east-west directions — exhibit large amplitude oscillations, especially at the beginning of the observation. These variations are consistent with the $\delta \rm{TEC}$ fluctuations seen in Figures~\ref{fig:dtec_antcs}–\ref{fig:dtec_antnorth}. The $p_0$ coefficient, related to the north-south gradient, shows fluctuations up to $\pm 5$ TECU, particularly during sunset. This behaviour aligns with known properties of medium-scale travelling ionospheric disturbances Medium-Scale Traveling Ionospheric Disturbance (MSTIDs) \citep{Her06}. Observation logs also suggest increased scintillation during this interval. Higher-order polynomial terms capture localized, finer-scale features more prominent at later times.

Smaller and more transient TEC gradient structures were observed in the middle portion of the observation (Figure~\ref{fig:tec_grad_pfit}). The north-south gradient showed more variability than the east-west component. For antennas on the GMRT's southern arm, the peak gradient amplitude reached approximately $\pm 0.01$ TECU/km. This behaviour is indicative of medium-scale travelling ionospheric disturbances (MSTIDs), which typically arise during the evening and nighttime \citep{Helmboldt12, Mangla22}. These small-scale features may reflect rippling in MSTID wavefronts, turbulent mixing from ion-neutral interactions, or localized enhancements, such as sporadic-E layers \citep{Coker09}. Due to the irregular layout of the central square antennas, resolving the directionality of TEC gradients was difficult in this region. Therefore, these antennas were excluded (Figure~\ref{fig:tec_grad_pfit}).

\subsection{Structure Function}
\label{sec:strf}

In this subsection, we characterize the spatial structure of the ionosphere using the \textit{phase structure function}, defined as
\begin{equation}
\Xi(r) = \left\langle \left(\phi(\vec{x}) - \phi(\vec{x} + \vec{r})\right)^2 \right\rangle,
\label{eq:phasestructure}
\end{equation}
where \( r \equiv |\vec{r}| \) is the separation distance, and \( \langle \cdot \rangle \) denotes the statistical expectation value.

The phase structure function is often modelled as a power law based on Kolmogorov’s theory of turbulence:
\begin{equation}
\Xi(r) = \left( \frac{r}{r_{\rm diff}} \right)^\beta,
\label{eq:Kolmogorov}
\end{equation}
where \( r_{\rm diff} \) is the \textit{diffractive scale} of the ionosphere. According to Kolmogorov’s theory for inertial-range turbulence, the power-law index is expected to be \( \beta = 5/3 \approx 1.67 \). A smaller diffractive scale corresponds to stronger phase fluctuations across the field of view, resulting in reduced temporal coherence of the signal \citep{Vedantham15}.

Deviations from the Kolmogorov scaling can occur due to non-turbulent ionospheric structures such as travelling ionospheric disturbances (TIDs) \citep{Vanvelthoven90} or field-aligned density ducts \citep{Loi15}. For example, wave-like TIDs typically yield slopes of \( \beta \sim 2 \).

\begin{figure}[H]
\centering
\includegraphics[width=\linewidth]{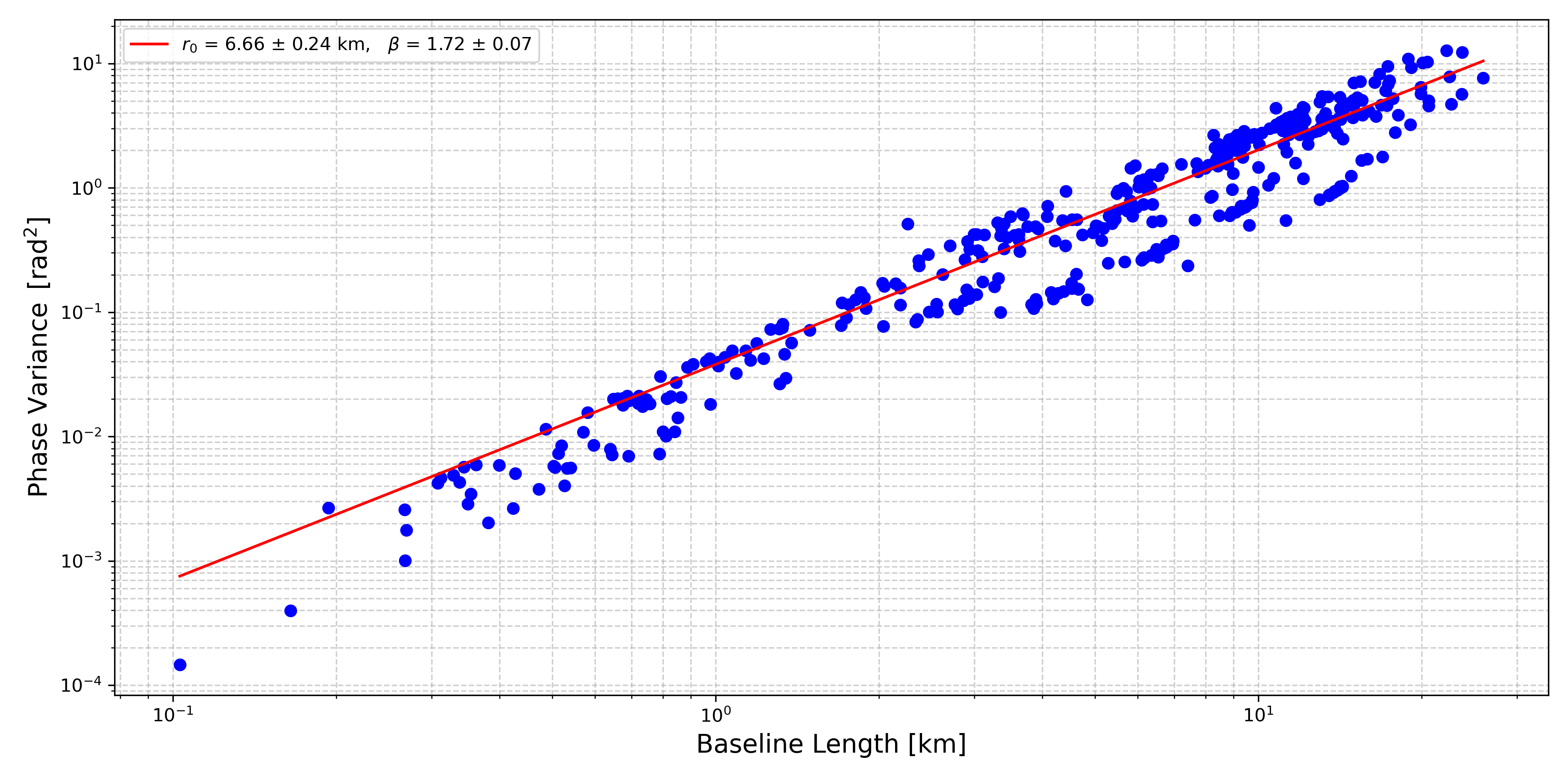}
\caption{Phase structure function of the ionosphere derived from differential TEC values, converted to phases at 600 MHz.}
\label{fig:sf}
\end{figure}

Figure~\ref{fig:sf} presents the measured phase structure function along with the best-fit model. We obtain a power-law index of \( \beta = 1.72 \pm 0.07 \), slightly steeper than the Kolmogorov value, suggesting the presence of additional ionospheric structures. The fitted diffractive scale is \( r_{\rm diff} = 6.66 \) km, which corresponds to the spatial scale over which the phase variance reaches 1~rad\(^2\).

Consistent with LOFAR observations by \citet{Mevius16}, we observe a band-like pattern in the structure function, indicative of anisotropic ionospheric irregularities. A full two-dimensional structure function analysis using uGMRT data would help determine whether these structures align with Earth's magnetic field \citep{Mevius16}. We leave this investigation to future work, where we plan to conduct a detailed 2D structure-function analysis across multiple uGMRT bands.

\section{Summary and Conclusion}
\label{sec:discussion}
We observed the bright flux calibrator 3C48 using uGMRT Band-4 data centred at 600 MHz, demonstrating the telescope’s unique ability to probe ionospheric activity at low latitudes. Our analysis shows that GMRT can measure differential TEC with sub-mTECU precision on 10-second timescales, offering significantly greater sensitivity to TEC fluctuations than GPS-based relative TEC measurements \citep{Her06}. The accuracy of these $\delta \rm{TEC}$ estimates could be further enhanced with improved sky and instrument modelling. Similar levels of precision have been achieved in mid-latitude studies with the VLA \citep{Helmboldt12} and LOFAR \citep{Mevius16}.

We find that for central square baselines with spatial scales $\leq$ 1.4 km, TEC fluctuations can be measured with a median absolute deviation (MAD) of 2.3 mTECU. In contrast, the precision drops by over an order of magnitude on longer baselines ($\geq$ 1.4 km). This decrease is likely due to small-scale ionospheric structures de-correlating over larger distances, which reduces signal coherence and limits sensitivity to differential TEC ($\delta \rm{TEC}$) variations. Factors such as system temperature, pointing accuracy, and variations in antenna beam shapes between central and remote antennas may also impact phase stability, further influencing the accuracy of $\delta \rm{TEC}$ estimates.

During our observations, we recorded significant fluctuations in the differential total electron content ($\delta {\rm TEC}$), with amplitudes increasing rapidly after local sunset at approximately 17:56 Indian Standard Time (IST). This timing aligns with the ionospheric transition from a photo-ionization driven regime to one dominated by recombination, a period known to foster enhanced plasma instabilities \citep{Dasgupta08}. Moreover, pre-reversal enhancement increases the upward movement of charged particles (plasma) in the ionosphere after sunset. This strengthens the Equatorial Ionization Anomaly (EIA), causing more electrons to accumulate in the two crest regions on either side of the magnetic equator. As radio signals pass through the resulting plasma bubbles, they become scattered, leading to amplitude and phase scintillations \citep{Balan18}.

We quantified the spatial variability of the ionosphere using the phase structure function and found that it follows a power-law behaviour with a slope \( \beta = 1.72 \pm 0.07 \), slightly steeper than the Kolmogorov prediction of \( 5/3 \). This suggests the presence of additional, possibly non-turbulent structures such as travelling ionospheric disturbances (TIDs) or field-aligned density ducts \citep{Loi15}. The derived diffractive scale of \( r_{\rm diff} = 6.66 \)~km indicates significant phase fluctuations across the field of view, with implications for signal coherence in low-frequency interferometric observations. The observed anisotropic features in the structure function further motivate a two-dimensional analysis to investigate potential alignment of ionospheric irregularities with Earth's magnetic field, which we defer to future work.

To evaluate the spatial coherence of ionospheric fluctuations across the GMRT array, we applied Principal Component Analysis to the time series of differential TEC ($\delta \rm{TEC}$) measured along each baseline. Interestingly, just five principal components explained nearly $100 \%$ of the total variance. This indicates that the ionospheric perturbations are highly structured and primarily governed by a small number of dominant temporal modes. Such a low-rank representation is consistent with a spatially coherent ionospheric screen, where fluctuations are strongly correlated across the array—even over relatively short baselines. This structure supports the application of reduced-order ionospheric models in calibration and points toward the presence of large-scale physical phenomena, such as travelling ionospheric disturbances or global TEC gradients. However, to robustly identify the nature of these fluctuations—such as their size, speed, and direction—and to distinguish them from other ionospheric processes, a more detailed spectral analysis of the time series is essential \citep{Helmboldt12b}.

The polynomial-based method described in Section \ref{sec:tec_grad} successfully captures the large-scale structure of the two-dimensional TEC gradients across the array. Although this technique lacks sensitivity to small-scale or rapidly varying features, it effectively models the broader, more dominant fluctuations—those with larger amplitudes and longer temporal or spatial scales—present in the TEC gradient field. Given that the $\delta \rm{TEC}$ measurements are accurate to about $1 \times 10^{-3}$ TECU 
and the average distance between antennas is around 3 km, this translates to a typical uncertainty in the estimated TEC gradient of roughly $\sim 3 \times 10^{-4}$ TECU per kilometre.

Finally, we conclude that the GMRT's geographic location makes it particularly advantageous for probing ionospheric behaviour, a key factor in enhancing the performance of low-frequency radio astronomy \citep{Intema09, Gasperin18}. If ionospheric systematics can be effectively captured using a small set of parameters, the number of free parameters in calibration can be significantly reduced—especially for observations with low signal-to-noise ratios. This approach paves the way for improved sensitivity and resolution in radio imaging \citep{Gasperin19}. Similar ionospheric studies will soon become feasible with upcoming next-generation instruments such as the SKA, expected to be operational in the coming years.

\authorcontributions{DB and AG coordinated the observational campaign, performed the radio data
analyses and interpretations, and led the paper writing. All the authors checked the paper and helped with the paper writing.}

\funding{All the authors wish to acknowledge funding provided under the SERB-SURE grant SUR/2022/000595 of the Science \& Engineering Research Board, a statutory body of the Department of Science \& Technology (DST), Government of India.}

\dataavailability{The observed data are publicly available after a proprietary period of 18 months through the GMRT data archive\footnote{\url{https://naps.ncra.tifr.res.in/goa/}} under the proposal code 47\_003.} 

\acknowledgments{We thank the staff of GMRT for making this observation possible. GMRT is run by the National Centre for Radio Astrophysics (NCRA) of the Tata Institute of Fundamental Research (TIFR). We thank all the reviewers whose comments helped us improve the overall quality of the paper. The authors utilized ChatGPT for AI-assisted copy editing and improving the manuscript's language. AG would like to thank IUCAA, Pune, for providing support through the associateship programme, including access to the computational facility at IUCAA.}

\conflictsofinterest{The authors declare no competing financial interests.} 

\appendixtitles{no} 
\appendixstart
\appendix
\section[\appendixname~\thesection]{}
\subsection[\appendixname~\thesubsection]{$\delta \mathrm{TEC}$ vs. Time for Central Square antennas (CASA flagging vs. AOFlagger)}

In this section, we compare the temporal variation of differential Total Electron Content ($\delta$TEC) as obtained using two different radio frequency interference (RFI) mitigation strategies: CASA's internal flagging algorithm and the AOFlagger tool. The focus is restricted to the GMRT's central square antennas, the arm antennas also show a similar behavior. The figure display the $\delta$TEC values obtained from each flagging method, allowing an assessment of how RFI mitigation influences ionospheric phase extraction. The red vertical dashed line marks the period during which scintillation was reported in the observation log.

\begin{figure}[H]
    \centering
    \includegraphics[width=\linewidth]{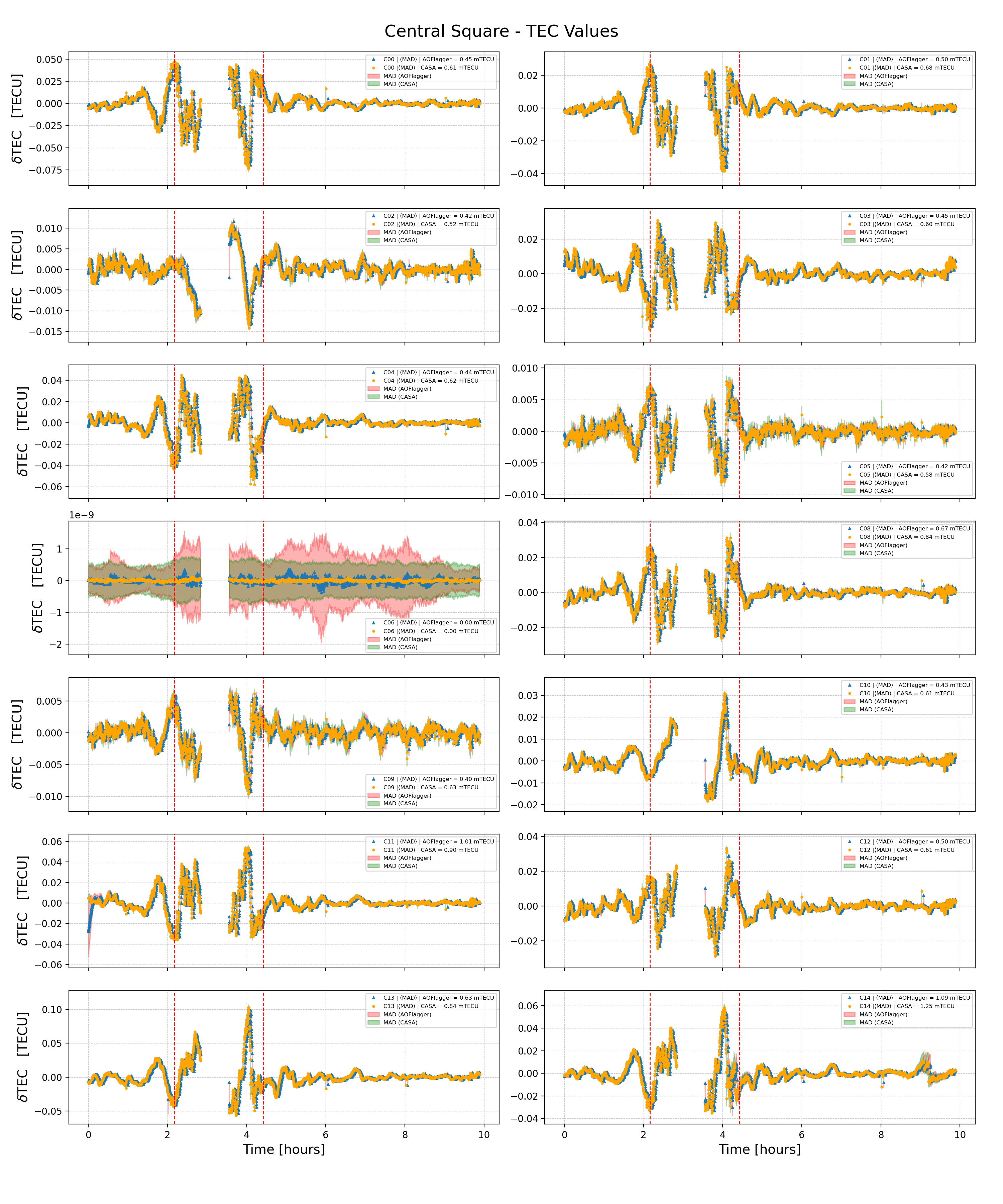}
    \caption{The blue triangles and orange circles illustrate the temporal variation of $\delta \rm{TEC}$ for the central square antennas. CASA flagging was applied for the blue data points, while AOFlagger was used for the orange ones. The red vertical dashed line indicates the time slots during which scintillation was reported in the GMRT observation log file.}
    \label{fig:tec_censq_casa_aoflag}
\end{figure}

\subsection[\appendixname~\thesubsection]{IRI-Predicted Peak Height and Slant TEC Scaling Factor}

This section presents model-based predictions of the ionospheric peak electron density height using the International Reference Ionosphere (IRI) model. The top panel of the figure shows how the peak height evolves over the course of the observation. This information is critical for determining the ionospheric pierce point geometry. In the bottom panel, we plot the corresponding slant-to-vertical TEC conversion factor, computed assuming a thin-shell ionospheric model at the predicted heights. 

\begin{figure}[H]
    \centering        
    \includegraphics[width=\linewidth, height=0.3\textheight]{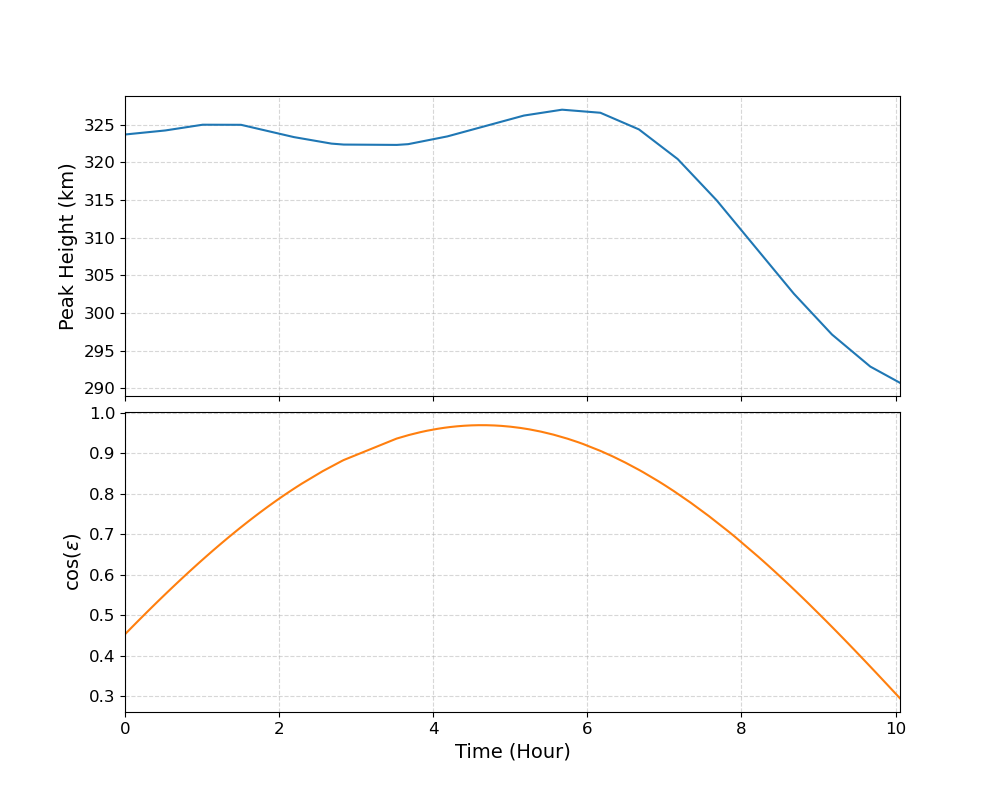}
    \caption{(Top) Temporal variation in the peak height of maximum electron density at the ionospheric pierce points, as predicted by the IRI model (\url {https://www.ionolab.org/iriplasonline/}). (Bottom) Associated scaling factor used to convert slant $\delta$TEC measurements to their vertical equivalents, assuming a thin-shell ionosphere positioned at the heights shown above.}
    \label{fig:ion_height_cosep}
\end{figure}

\subsection[\appendixname~\thesubsection]{Projected TEC gradient}

This section shows the TEC gradient projected along the GMRT arms, estimated from polynomial fits to $\delta$TEC time series data between antenna pairs. Gradients are plotted for the north-south (blue) and east-west (orange) directions, revealing the anisotropic structure of ionospheric variations.

\begin{figure}[H]
    \centering
    \includegraphics[width=\linewidth, height=0.4\textheight]{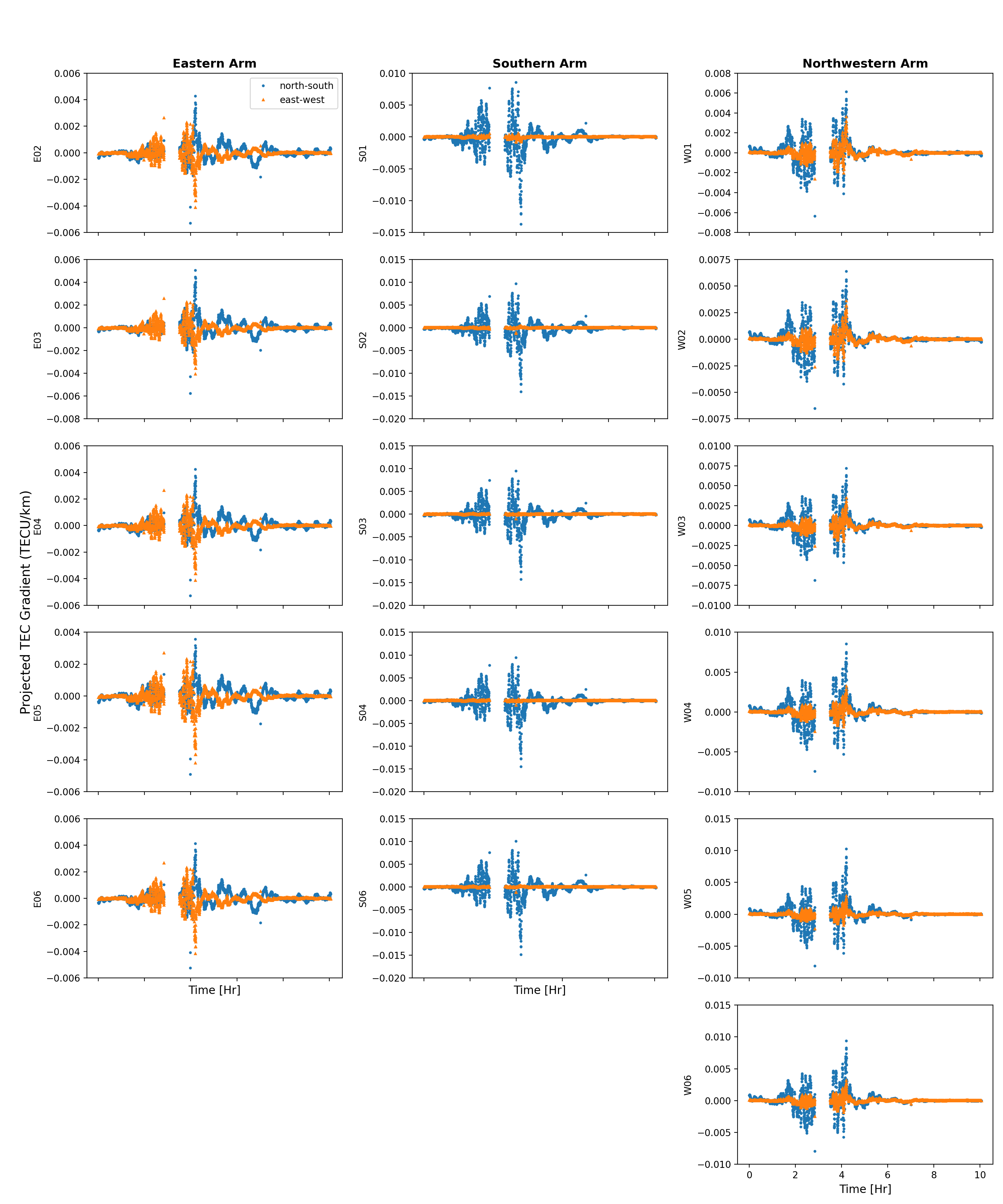}
    \caption{This figure shows the projected TEC gradients estimated from the polynomial coefficients for the GMRT arm antennas along north-south (blue circles), and east-west (orange triangles) directions.}
    \label{fig:tec_grad_pfit}
\end{figure}

\subsection[\appendixname~\thesubsection]{Absolute vertical TEC from IGS models}
The ionospheric Total Electron Content (TEC) estimates on an hourly basis from different IONEX data centers, we used the Spinifex library (\url{https://git.astron.nl/RD/spinifex}) to extract vertical TEC (vTEC) values along the line of sight to the radio source 3C48, as seen from the GMRT site. For this estimate, we adopted a single-layer ionospheric model at a height of 300 km.

\begin{figure}[H]
    \centering
    \includegraphics[width=\linewidth, height=0.4\textheight]
    {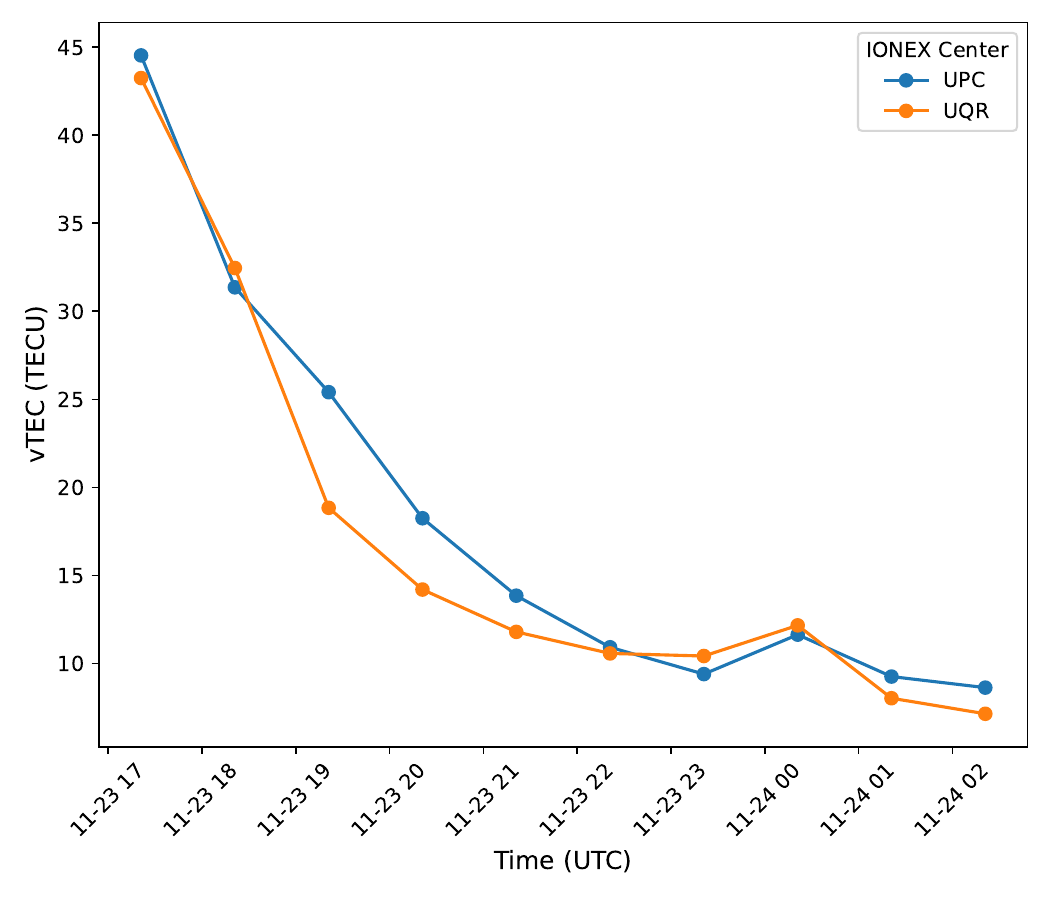}
    \caption{This figure presents hourly estimates of the ionospheric Total Electron Content (TEC) from various IGS models during our observation period. The overall trend aligns with our differential TEC measurements, showing greater ionospheric variability at the beginning of the observation, which gradually decreases toward the end.}
    \label{fig:vtec_igs}
\end{figure}

\reftitle{References}

\bibliography{ref}

\PublishersNote{}

\end{document}